# Improvement in granularity of NdFeAsO$_{0.8}$F$_{0.2}$ superconductor through Ag$_x$ doping (x = 0.0 – 0.3)

Poonam Rani[1,2], A.K. Hafiz[2] and V. P. S. Awana[1,*]

[1]National Physical Laboratory (CSIR) Dr. K. S. Krishnan Road, New Delhi-110012, India

[2] Department of Physics, Jamia Millia Islamia, New Delhi-110025, India

ABSTRACT

We report the impact of silver addition on granularity of NdFeAsO$_{0.8}$F$_{0.2}$ superconductor. The *ac* susceptibility and electrical resistivity under magnetic field are measured to study the improvement in weak links of NdFeAsO$_{0.8}$F$_{0.2}$ with addition of Ag$_x$ (x = 0.0 – 0.3). The Ag free NdFeAsO$_{0.8}$F$_{0.2}$ compound shows superconductivity at around 51.8K. Typical two step superconducting transitions due to the inter and intra grain contributions, induced from the combined effect of superconducting grains and the inter-granular weak-coupled medium respectively are clearly seen in susceptibility [$\chi(T)$] plots. In comparison to the pure NdFeAsO$_{0.8}$F$_{0.2}$ compound, the coupling between the superconducting grains is significantly improved for 20wt% silver doped sample, and the same is deteriorated for higher Ag content i.e., for 30wt% Ag sample. The magneto transport measurements $\rho(T)H$ of polycrystalline 20wt% Ag doped NdFeAsO$_{0.8}$F$_{0.2}$, exhibited the upper critical filed [$H_{c2}(0)$] of up to 334Tesla, which is slightly higher than the one observed for pure NdFeAsO$_{0.8}$F$_{0.2}$. The flux flow activation energy ($U_0/k_B$) varies from 7143.38K to 454.77K with magnetic field ranging from 0Tesla to 14Tesla for 20wt% Ag doped NdFeAsO$_{0.8}$F$_{0.2}$. In this investigation, our results show that limited addition of Ag improves the granular coupling of superconducting grains of NdFeAsO$_{0.8}$F$_{0.2}$ compound.

*: Corresponding Author: e-mail: awana@mail.nplindia.org
Phone/Fax – 0091-11-45609357/9310: Home Page: awanavps.webs.com

INTRODUCTION

The exciting research of Iron based superconducting compounds RE($O_{1-x}F_x$)FeAs with RE = rare earth (1111 phase), AEFe$_2$As$_2$ with AE = alkali or alkali earth (122 phase), AFeAs with A = alkali metal (111 phase) and FeSe (11 phase) [1-5] has attracted tremendous interest after being discovered by Kamihara group in 2008 [1]. The RE($O_{1-x}F_x$)FeAs (1111 phase) for the different rare earths (La, Ce, Pr, Gd, Sm and Nd) show varying superconducting transition temperature ($T_c$) as replacing lanthanum with heavier rare-earth elements resulted in an increase of $T_c$ from 26K to above 55K[1, 6-11]. Interestingly, these compounds are similar to the High $T_c$ cuprate superconductors (HTSc cuprates) in terms of their layered structure, low carrier density and magnetically ordered parent phase [12]. At the same time, Fe based superconductors are different from HTSc cuperates in terms of their very high upper critical fields, and low value of crystallographic anisotropy, which makes them more favourable for practical applications [13]. Like HTSc cuprates, the granularity in pnictide superconductors is one of the important factors that limit their practical applications. In this regards, the investigations on critical current density ($J_c$) along with short coherence length ($\xi$) of 1111 polycrystalline samples have already shown significant evidence for the weak granular coupling between the superconducting grains of these new superconductors [14, 15]. Many researchers have devoted their research in improving the weak links coupling, but still in preliminary stage due to the lake of familiarity with origin and mechanism behind the granularity in these systems. In this regards, the general feeling is that due to the small value of the coherence length, the grain boundaries as well as the defects even of small dimensions act as weak links, giving rise to flux penetration at low magnetic fields and hence the ensuing depression of the critical current density [16-20]. Also, it has not yet been clear as if such effects are due to intrinsic properties of the oxy-pnictides or are related to specific properties of the investigated samples. The *ac* susceptibility and resistivity measurements under magnetic field are suitable tools for the investigation of the granular nature of the superconductors. One can very clearly observe the nonlinear effect due to inter and intra granular coupling of grains. For example the *ac* susceptibility at low DC magnetic fields as a function of the *ac*-field amplitude allows highlighting the possible nonlinear effects related to critical state in the inter-grain region [21].

As far as the improving of grain boundaries connectivity is concerned, a lot of work has already been done in case of cuprate superconductors and the addition of Ag has proven to be very effective [22-26]. Infact, to draw wires/tapes of high $T_c$ cuprates for their practical applications, researchers have drawn silver sheathed wires and tapes of them for their high current applications [27-29]. It is known that Ag does not have any deteriorating effect on superconductivity of HTSc cuprates, primarily because Ag gets distributed at the grain boundaries instead of being substituted in the parent superconducting unit cell of these compounds [22-29]. This way Silver addition has resulted in improving the critical current of high Tc cuprates in a big way. This prompted us to try Silver addition in Nd based 50K Fe pnictide NdFeAsO$_{0.8}$F$_{0.2}$ superconductor and to check if there is any positive impact of the same. Infact there are a couple of studies already in literature related to the addition of Silver in Fe pnictide superconductors, but yet in a sense are of primitive nature [30-32] and a kind of first notes on this topic.

Here we investigate the NdFeAsO$_{0.8}$F$_{0.2}$ superconductor added with different atomic weight percent of Ag for detailed magneto-transport and *ac* susceptibility measurements. The magnetic and transport properties have been measured through PPMS (Quantum design) under magnetic field up to 14Tesla. The morphology and the grain connectivity of the all samples have been seen from the Scanning Electron Microscopy (SEM). It is found that limited addition of Ag improves the granular coupling of Fe pnictide NdFeAsO$_{0.8}$F$_{0.2}$ 50K superconductor.

EXPERIMENTAL

A series of samples NdFeAsO$_{0.8}$F$_{0.2}$-Ag$_x$ (x = 0.0, 0.05, 0.10, 0.20, 0.30) were synthesized through single step solid-state reaction route via vacuum encapsulation technique. High purity (~99.9%) powders of Nd, As, Fe$_2$O$_3$, Fe, and FeF$_3$ in their stoichiometric ratios are properly weighed, mixed and ground thoroughly using mortar and pestle in presence of high purity Ar atmosphere in glove box. The Humidity and Oxygen content in the glove box are maintained at less than 1ppm. The mixed powders were palletized and vacuum-sealed (10$^{-4}$Torr) in a quartz tube. Then sealed pellet was placed in box furnace for sintering under the heat treatment, 550ºC for 12h, 750ºC for 12h, 950ºC for 12h and then at 1140ºC for 12h in continuum with slow heating rate. Then furnace is cooled slowly down to room temperature. Sintered sample is obtained in the form of black powders after breaking the sealed quartz tubes. Phase purity has been checked through XRD. Then finally the sample was ground after mixing the known but varying Ag (0-30wt%) metal powder, pelletized and then sealed in a quartz tube. Sealed samples are finally treated at 1150ºC for 12h to obtain hard pellets of samples. Subsequently, the furnace is allowed to cool naturally.

The synthesized samples are characterized through room temperature X-ray diffraction using Rigaku X-ray diffractometer with Cu $K_\alpha$ radiation. The magnetic and resistivity measurements under magnetic field were carried out by a conventional four-probe method on a quantum design Physical Property Measurement System (PPMS) with fields up to 14Tesla.

RESULTS AND DISCUISSION

Figure 1 shows the Rietveld fitted X- ray diffraction patterns of 0wt%, 5wt%, 10wt% and 20wt% of Ag doped NdFeAsO$_{0.8}$F$_{0.2}$ samples. Fitted data show that the studied samples are mostly single phase. All the samples are crystallized in tetragonal ZrCuSiAs type structure having space group P4/nmm. Also minute impurity of NdOF is observed in all samples, which have been marked as * in Fig 1. We did not observe any peak of added Ag till 20wt% doping but for further doping (30wt%) some peaks of Ag are observed (not shown in XRD pattern). This is possible, because till 20wt% Ag doping the same might be getting uniformly distributed at nano-metric grain boundaries. On the other hand for higher concentrations (30wt%) of Ag the clustering might be taking place and as result the same could be seen in XRD. The lattice parameters obtained from the XRD for the un-doped and the Ag-doped samples are included in Table I. Compared with the pristine sample, the lattice parameter 'c' is found to increase slightly but no change is observed in 'a' parameter with 5wt% Ag doping. On further doping of Ag (20wt%) slight decrease in both a and c parameter is seen. But for the sample with Ag (30wt%) both a and c values increase slighly.

The temperature dependent *ac* magnetic susceptibility in both zero fields cooled (ZFC) and field cooled (FC) situations are shown in Figure 2. The applied magnetic field is 50Oe. The pure NdFeAsO$_{0.8}$F$_{0.2}$ compound exhibits superconducting transition temperature ($T_c$) at below 48.2K in terms of clear diamagnetic transition. The $T_c$ decreases continuously to 47K, 45.8K and 44.3K for 5wt%, 10wt% and 20wt% Ag doped NdFeAsO$_{0.8}$F$_{0.2}$ samples. Both FC and ZFC are nearly saturated at around 35K in case of pure, 5wt% and 10wt% Ag doped samples. On the other hand, the 20wt% Ag doped NdFeAsO$_{0.8}$F$_{0.2}$ behaves differently in FC and ZFC situations exhibiting two step diamagnetic transitions.

Figure 3 (a) shows the temperature dependent real part of *ac* susceptibility ($\chi'$) for pure NdFeAsO$_{0.8}$F$_{0.2}$ sample at different ac field amplitudes of 1Oe - 17Oe and at a fixed frequency of 333Hz. The measurement is performed at zero *dc* fields. Worth mentioning is the fact, that for achieving the zero *dc* field at 14 Tesla PPMS, various +ve and −ve ramping of dc field are necessary. The diamagnetic transition is observed at 47K with *ac* driven field amplitude of 1Oe. Intensity of transition increases with increasing the field amplitudes. At low field amplitude of up to 5Oe the transition is found to be single and nearly saturating, as complete shielding of superconducting grains takes place below $T_c$. As field amplitude is increased from above 5Oe to 17Oe a non saturating behaviour is observed, indicating the response of weakly coupled superconducting grains at low temperatures i.e., at below $T_c$ [34]. This is an indication of the granular nature of the studied polycrystalline NdFeAsO$_{0.8}$F$_{0.2}$ superconductor, which will be clearer when we study the imaginary part of *ac* magnetic susceptibility in next section.

The imaginary part of *ac* susceptibility ($\chi''$) for NdFeAsO$_{0.8}$F$_{0.2}$ is shown in Figure 3 (b). It is clear that only inter granular peak is observed close to 35K at 1Oe *ac* field amplitude. At higher *ac* field of above 5Oe though the $\chi''$ data show distinct intra-granular peak at around 45K, i.e., close to $T_c$, but the inter-granular peak gets broadened and is not complete. Interestingly, the inter-granular peak is observed for lower amplitudes only. The peak near to $T_c$ arises, when the *ac* field penetrates to the centre of grains, and the peak at lower temperature corresponds to the complete penetration of the *ac* field in to the centre of the sample [35,36]. The appearance of a kink at ~45K at higher amplitudes clearly demonstrates the presence of intra-grain superconductivity of the sample. This means at these amplitudes the inter-grain superconductivity due to weak coupling of individual grains is nearly disappeared and that of intra-grain (individual grain) is appeared. Though the intra-grain peak position remains nearly the same, but the inter-grain peak shifts towards the lower temperature with increasing *ac* field from 1–17Oe. Namely, the inter-granular peak temperature ($T_p$) decreases from 35K to 10K, when the applied *ac* field ($H_{ac}$) amplitude is increased from 1Oe to 9Oe. This gives rise to a dTp/dH$_{ac}$ of ~3.12K/Oe. Interestingly, this value of dTp/dH$_{ac}$ of ~3K/Oe for NdFeAsO$_{0.8}$F$_{0.2}$ superconductor is more than double to that as observed for bulk YBa$_2$Cu$_3$O$_7$ high $T_c$ cuprate, for which the same is around 1.3K/Oe [37]. In general terms, the relative shift of inter-granular peak temperature ($T_p$) with *ac* amplitude is a direct measure of the superconducting grains coupling in a bulk superconductor. Lower is the dT$_c$/dH$_{ac}$, the stronger is the coupling between the superconducting grains. This shows that the coupling of superconducting grains of $NdFeAsO_{0.8}F_{0.2}$ is weaker in comparison to $YBa_2Cu_3O_7$ [37]. The *ac* susceptibility ($\chi^{//}$) for $NdFeAsO_{0.8}F_{0.2}$ for further higher (>9Oe) *ac* field amplitudes shows that the peak formation remains incomplete down to 10K, see (Fig. 3b). This again shows that the studied bulk polycrystalline for $NdFeAsO_{0.8}F_{0.2}$ superconductor is highly granular in nature and the inter-granular coupling is weak.

Figure 4 (a) and (b) shows the real and imaginary part of temperature dependence *ac* susceptibility for the 5wt% Ag doped $NdFeAsO_{0.8}F_{0.2}$ sample. Qualitatively it looks similar to that as in the case of pristine sample being shown in Fig. 3 (a) and (b). At lower *ac* fields single transition is found, but at higher fields, clear two transitions in superconducting region i.e. below 45K are seen. The first one (43K) being intra-grain and the second one (36K) is due to inter-grain superconducting transitions. Also, though the intra-grain transition remains almost unchanged at around 43K, the inter grain one shifts to lower temperatures with increase in *ac* amplitude from 1Oe to 17Oe. Although the *ac* susceptibility results of 5wt% Ag doped sample looks very similar to $NdFeAsO_{0.8}F_{0.2}$, but the quantitative point of view makes them different from each other. For example the inter-granular peak is sharper in case of 5wt% Ag doped sample. Also the relative change in inter-grain peak position ($T_p$) with *ac* fields is comparatively lower in case of 5wt% Ag doped sample than the pure $NdFeAsO_{0.8}F_{0.2}$ sample. In case of 5wt% Ag doped $NdFeAsO_{0.8}F_{0.2}$ sample, the $T_p$ value decreases from 36K (1Oe) to around 10K (17Oe). This gives rise to a $dTp/dH_{ac}$ of ~1.6K/Oe. Interestingly, this value is nearly half to that as observed for pristine $NdFeAsO_{0.8}F_{0.2}$ sample. This shows that superconducting grains of 5wt% Ag doped $NdFeAsO_{0.8}F_{0.2}$ sample are better coupled than the pristine $NdFeAsO_{0.8}F_{0.2}$.

The result of real and imaginary parts of *ac* susceptibility of 10wt% Ag doped $NdFeAsO_{0.8}F_{0.2}$ sample are displayed in Figures 5 (a) and 5 (b). As in the case of pure and 5wt% Ag doped $NdFeAsO_{0.8}F_{0.2}$ samples, two clear transition peaks (intra and inter) appear only with higher *ac* field amplitudes. For smaller (<5Oe) *ac* field amplitudes, the intra granular superconducting peak is not seen. Further, once appeared, the intra granular peak position is nearly invariant for all higher *ac* field amplitudes and the inter-granular peak shifts towards the lower temperature. The intra-grain peak near $T_c$ appears at around 43K, and its position is almost independent of the *ac* field amplitudes. Similar to other samples the inter-grain peak temperature ($T_p$) shifts towards lower temperature with increase of the *ac* field amplitudes. Quantitatively, the $T_p$ value shifts from 37K at 1Oe to 13K at 17Oe. This gives rise to a dTp/dH$_{ac}$ of ~1.5K/Oe. Seemingly, there is not much difference in superconducting performance of 5 and 10wt% Ag doped NdFeAsO$_{0.8}$F$_{0.2}$ samples.

Further, to observe the effect of higher silver on the granular coupling of NdFeAsO$_{0.8}$F$_{0.2}$, the real and imaginary part of *ac* susceptibility for the 20wt% Ag doped sample are shown in the Figures 6 (a) and 6 (b). In the real part, the superconducting transition is seen at 44K, and is relatively sharper in comparison to the pure and lower concentration of Ag added samples. The shape of the transition and the magnitude of the diamagnetic signal prove the improved superconductivity of this sample. It is clearer from imaginary part as well which is the replica of real part. Interestingly, the inter and intra granular peaks seems to approach each other, as if the compound has become mono grain with completely fused grains. Or otherwise one can say that this amount of silver is enough to suppress inter granular peak. Surely enough, the intra granular peak is more prominent in this case. Here only the formation of intra granular peak is complete and inter granular peak is almost disappeared. A little shift in the intra granular peak has also been observed i.e. from 42K (1Oe) to 37K (17Oe). Occurrence of intra granular peak confirms the presence of strong coupling between superconducting grains with least grain boundaries. It indicates that the inter-granular coupling is improved significantly in 20wt% Ag doped sample in comparison to pristine sample.

The real and imaginary parts of *ac* susceptibility of NdFeAsO$_{0.8}$F$_{0.2}$ + 30wt% Ag sample are shown in Figures 7 (a) and 7 (b). In this case again both the intra and inter granular transitions are seen. Considering the imaginary part, it is clear that the inter granular peak formation is not complete due to the similar behaviour of grains as did in the previous cases of pure, 5wt% and 10wt% of Ag doped samples. The T$_p$ value at 1Oe is 35K and at 17Oe is around 13K. This gives rise to a dTp/dH$_{ac}$ of ~1.38K/Oe. It seems the excess Ag in case of 30wt% Ag doped sample lies at grain boundaries in bulk chunks instead of possible uniform distribution as in case of 20wt% Ag sample. Clearly, the 30wt% Ag doped sample is much inferior in its superconducting performance in comparison to the 20wt% Ag doped sample.

Figure 8 shows the temperature dependent resistivity of NdFeAsO$_{0.8}$F$_{0.2}$, 5wt%, 10wt% and 20wt% Ag doped NdFeasO$_{0.8}$F$_{0.2}$. The addition of silver decreases the critical temperature T$_c$ from 51.8K for pure NdFeAsO$_{0.8}$F$_{0.2}$ to 47.2K for 20wt% Ag doped NdFeAsO$_{0.8}$F$_{0.2}$. All the samples show metallic behaviour above superconducting transition temperature. Despite small (~4K) decrease of $T_c$, all samples have similar non-Fermi temperature dependence of the resistance in the normal state. Specifically, it shows a remarkable linearity of the ρ (T) curves just above the critical temperature $T_c$. The values of $T_c$, $\Delta T_c$ (transition width = $T_c^{onset}$ - $T_c^{\rho=0}$), normal state resistivity ($\rho_{300}$) and residual resistivity ratio (RRR) obtained from ρ-T are tabulated in Table II. The calculated values of $\Delta T_c$ displayed in Table II are comparable with other pnictides [3, 9]. The (RRR = $\rho_{300}/\rho_{60}$) value of the 20wt% Ag doped sample is less in comparison to other measured samples. The table also exhibits a slight decrease in $\Delta T_c$ with increasing Ag concentration. It is to be noted that the pure sample has the maximum $T_c$, highest normal state resistivity and the highest RRR value. The variation of $T_c^{onset}$ with Ag concentration is shown in inset of Fig. 8.

Figure 9 (a-d) shows the temperature dependence of normalised resistivity ($\rho/\rho_{60}$) of pure $NdFeAsO_{0.8}F_{0.2}$, 5wt% Ag doped, 10wt% Ag doped and 20wt% Ag doped $NdFeAsO_{0.8}F_{0.2}$ respectively under magnetic field from 0 to 14Tesla. The field is applied perpendicular to the direction of current flow. At zero magnetic field, the $T_c(\rho=0)$ of pure sample (Fig. 9a) is around 46.8K and the onset transition temperature $T_c^{onset}$ is at 51.8K . The $T_c^{onset}$ remains nearly unaffected with applied magnetic field while the offset temperature ($T_c^{\rho=0}$) shifts towards the lower temperatures and decreased to below 30.3K on applying 14Tesla magnetic field. Shift in the offset transition with magnetic field leads to the broadening of transition width. Although the $T_c^{onset}$ remain nearly invariant, but a noticeable shift has been observed $T_c^{\rho=0}$ for of the all the samples. For pure sample the offset shifts from 46.8K (0Tesla) to 30.3K (14Tesla) hence the rate of decrease of offset transition temperature with applied field for pure sample is around 1.18K/Tesla {$dT_c/dH$ = (46.8K-30.3K)/(14-0Tesla)}. Doping of silver improves the transition width as well as broadness of the offset transition temperature under magnetic field. Though, onset of 5wt% Ag doped $NdFeAsO_{0.8}F_{0.2}$ without field is 48.8K, which is slightly lower as compared to the pure sample (51.8K) but the offset transition temperature ($T_c^{\rho=0}$) under magnetic field is yet comparable to that as in case of pure sample, see Fig. 9(b). At 0 fields the offset transition temperature is 43.8K, which decreases to 27.6K at 14Tesla field. The rate of decrease of offset transition temperature ($dT_c/dH$) for 5wt% Ag doped sample is 1.16K/Tesla, which is slightly smaller than the one (1.18K/Tesla) as observed for pure (without Ag addition) sample.

On further increasing the concentration of silver not only the broadness of transition becomes less but also the rate of decrease of offset transition temperature also decreases with increase in magnetic field. For 10wt% Ag doped sample, the onset at 0 field is around 49.3K (Fig. 9c) and the same is at 47.2K (Fig. 9d) for 20wt% Ag doped sample. Although a decrease in onset is found in case of 20wt% Ag added sample but on seeing the broadness of offset with magnetic field one can find a great improvement in the granular coupling. The shift observed in offset transition temperature from 44.8K (0Tesla) to 30 K (14Tesla) for 10wt% Ag doped sample, whereas for 20wt% Ag doped sample the same shifts from 43.4K (0Tesla) to 29.5K (14Tesla). The rate of shift in the offset transition temperature with applied magnetic field for 10 and 20wt% Ag doped samples are around 1.06K/Tesla and 0.99K/Tesla respectively, which are far less in comparison to the pure sample, for which the same is 1.18K/Tesla (Fig. 9a). This shows that Ag has significantly improved the superconducting performance under magnetic field for the $NdFeAsO_{0.8}F_{0.2}$ superconductor.

Figure 10 presents the upper critical field [$H_{c2}(T)$] at zero temperature as calculated by extrapolating the data using Ginzburge-Landau (GL) equation. The $H_{c2}(T)$ is obtained using 90% criteria of $\rho_N$ (normal state resistivity) i.e. where resistivity becomes 90% of its normal state value. The Ginzburge-Landau equation is

$$H_{c2}(T) = H_{c2}(0) x [\frac{(1-t^2)}{(1+t^2)}]$$

Where $t = T/T_c$ is the reduced temperature and $H_{c2}(0)$ is the upper critical field at zero temperature. The $H_{c2}$ versus temperature plot finds that the $H_{c2}$ value of pure $NdAsO_{0.8}F_{0.2}$ is around 318Tesla, which is decreased slightly for up to 10% Ag doping. In case of 20% Ag doped sample, the $H_{c2}$ improves and the value obtained is 334Tesla.

Figures 11 (a-d) show the temperature derivative of resistivity (d$\rho$/dT) versus temperature for the $NdFeAsO_{0.8}F_{0.2}$ and 5wt% to 20wt% Ag doped $NdFeAsO_{0.8}F_{0.2}$ samples under magnetic fields of up to 14Tesla. In zero fields, a sharp peak is seen due to the good percolation path of superconducting grains indicating good coupling between the grains. Narrower is the peak better is coupling of the grains. In fact with Ag addition the d$\rho$/dT vs T plot peak becomes narrow and hence confirms the improved coupling of superconducting grains with silver doping. On applying the field the transition peak becomes broader with a shift towards the lower temperature. The peak intensity also decreases with increasing applied magnetic field. The broadening of peak under magnetic field suggests that the onset part is less affected by the magnetic field as compared to the offset part. This happens due to the thermally activated creeps of vortices [37, 38]. This interpretation is based on the fact that, for the low-resistivity region, the resistivity is caused by the creep of vortices so that the ρ(T) dependences are thermally activated which are usually described by an Arrhenius equation

$$\rho(B, T) = \rho_0 \exp[-\frac{U_0}{k_B T}]$$

Here, $U_0$ is the thermally activation flux-flow (TAFF) energy, which can be obtained from the slope of the linear part of an Arrhenius plot, $\rho_0$ is a field independent pre-exponential factor, and $k_B$ is Boltzmann's constant. The best fitted ln ρ vs. $T^{-1}$ plots to the experimental data being shown in Figure 12, yields the values of the activation energy ranging from $U_0/k_B$ = 4072.6K and 419.5K for pure $NdFeAsO_{0.8}F_{0.2}$ in the field range of 0-13Tesla, which is comparable with the other reported results on Fe pnictide superconductors [39, 40]. The $U_0/k_B$ are 4063.4K to 477.1K for 5wt% Ag, 4590.9K to 501K for 10wt% Ag and 7143.4 K to 454.8K for 20wt% Ag doped samples are determined in the field of up to 14Tesla. The variation of activation energy with magnetic field for various samples is shown in Figure 13. It can be seen that the $U_0$ value of $NdFeAsO_{0.8}F_{0.2}$ at lower fields decreases weakly and scales to ~$B^{-0.28}$, and then decreases abruptly as $B^{-0.73}$ at higher magnetic fields. Similar trend is found in the case of Ag doped samples. For 20wt% Ag doped $NdFeAsO_{0.8}F_{0.2}$ slope at lower field shows a slight increase in comparison to pure sample, being calculated to be as $B^{-0.34}$ while at higher fields the same is calculated to be $B^{-0.68}$.

The microstructure of polished pallets of pure $NdFeAsO_{0.8}F_{0.2}$ and Ag-doped $NdFeAsO_{0.8}F_{0.2}$ were investigated using SEM. Figures 14(a), 14 (b), 14 (c) show the SEM images of pure $NdFeAsO_{0.8}F_{0.2}$, 10wt% Ag doped and 20wt% of Ag doped $NdFeAsO_{0.8}F_{0.2}$ samples respectively. Plated type grains are found. The Ac susceptibility can be correlated with the microstructure of corresponding sample. 10wt% Ag sample seems to more porous than other samples. Porosity almost disappears in 20wt% Ag doped sample and grains are appeared to stick with each other. Comparatively, better granular coupling is appeared in 20wt% Ag doped sample.

**Conclusion**

From the above results it is clear that 20wt% of Ag doping significantly improves the superconducting performance of the pure $NdFeAsO_{0.8}F_{0.2}$ sample. It is seen that though the $T_c$ decreases marginally, the superconducting performance under magnetic field is improved significantly with Ag doping of up to 20wt%. Clearly, like in high $T_c$ cuprates, the Ag addition in limited quantity improves granular coupling of Fe pnictide superconductor. This study brings out the fact that the wires and tapes of Fe pnictide superconductors could be successfully fabricated by Ag cladding powder in tube method.

**Acknowledgements**

Poonam Rani would like to thank the University for the Support of university fellowship to pursue her Ph.D. This research at NPL is supported by DAE-SRC outstanding researcher grant no. GAP 123332.

**Figure Captions**

**Figure 1:** Retvield fitted room temperature observed X-ray diffraction patterns for representative samples of NdFeAsO$_{0.8}$F$_{0.2}$, 5wt%, 10wt% and 20wt% Ag doped NdFeAsO$_{0.8}$F$_{0.2}$.

**Figure 2:** Variation of DC volume susceptibility with temperature of NdFeAsO0.8F0.2 with variable Ag content.

**Figure 3:** (a) and (b) Real and imaginary part of Ac susceptibility of Pure NdFeAsO$_{0.8}$F$_{0.2}$ as a function of temperature for different field amplitude varies from 0Oe to 17Oe.

**Figure 4:** (a) and (b) Real and imaginary part of Ac susceptibility of 5wt% Ag doped NdFeAsO$_{0.8}$F$_{0.2}$ as a function of temperature for different amplitude varies from 0Oe to 17Oe.

**Figure 5:** (a) and (b) Real and imaginary part of Ac susceptibility of 10wt% Ag doped NdFeAsO$_{0.8}$F$_{0.2}$ as a function of temperature for different amplitude varies from 0Oe to 17Oe.

**Figure 6:** (a) and (b) Real and imaginary part of Ac susceptibility of 20wt% Ag doped NdFeAsO$_{0.8}$F$_{0.2}$ as a function of temperature for different amplitude varies from 0Oe to 17Oe.

**Figure 7:** (a) and (b) Real and imaginary part of Ac susceptibility of 30wt% Ag doped NdFeAsO$_{0.8}$F$_{0.2}$ as a function of temperature for different amplitude varies from 0Oe to 17Oe.

**Figure 8:** Resistivity behaviour with temperature variation ρ(T) of representative samples of NdFeAsO$_{0.8}$F$_{0.2}$ with variable Ag contents at zero field. Inset shows the electronic phase diagram of Ag-doped NdFeAsO$_{0.8}$F$_{0.2}$.

**Figure 9:** Resistivity behaviour in presence of magnetic field ρ(T)H upto 14T for pure NdFeAsO$_{0.8}$F$_{0.2}$, 5wt% Ag doped, 10wt% Ag doped and 20wt% Ag doped NdFeAsO$_{0.8}$F$_{0.2}$ samples respectively.

**Figure 10:** The Ginzburge Landau (GL) fitted variation of upper the upper critical field (H$_{c2}$) with temperature for NdFeAsO$_{0.8}$F$_{0.2}$ and Ag doped samples respectively at 90% criteria.

**Figure 11:** Temperature derivative of normalized resistivity for NdFeAsO$_{0.8}$F$_{0.2}$, 5wt%, 10wt% and 20wt% Ag doped NdFeAsO$_{0.8}$F$_{0.2}$ samples respectively.

**Figure 12:** Fitted Arrhenius plot of resistivity for NdFeAsO$_{0.8}$F$_{0.2}$, 5wt%, 10wt% and 20wt% Ag doped NdFeAsO$_{0.8}$F$_{0.2}$ samples respectively.

**Figure 13:** $U_0$ dependence of magnetic field of $NdFeAsO_{0.8}F_{0.2}$ with variable Ag contents.

**Figure 14**: High magnification SEM micrographs for the following samples (a) pure $NdFeAsO_{0.8}F_{0.2}$; (b) 10wt% Ag added; and (c) 20wt% Ag added $NdFeAsO_{0.8}F_{0.2}$.

**Table 1:** Reitveld refined lattice parameters with corresponding transition temperature.

| Compound | a (Å) | c (Å) | Vol. (Å$^3$) | $\chi^2$ | $T_c^{onset}$(K) |
|---|---|---|---|---|---|
| NdFeAsO$_{0.8}$F$_{0.2}$ | 3.971(1) | 8.571(3) | 135.17(7) | 2.54 | 51.80 |
| NdFeAsO$_{0.8}$F$_{0.2}$ + 5% Ag | 3.971(1) | 8.572(2) | 135.21(6) | 2.60 | 48.87 |
| NdFeAsO$_{0.8}$F$_{0.2}$ + 10% Ag | 3.960(1) | 8.544(3) | 134.02(8) | 2.85 | 49.34 |
| NdFeAsO$_{0.8}$F$_{0.2}$ + 20% Ag | 3.951(2) | 8.523(4) | 133.05(10) | 3.08 | 47.22 |
| NdFeAsO$_{0.8}$F$_{0.2}$ + 30% Ag | 3.964(1) | 8.549(4) | 134.38(9) | 3.95 | 43K-mag |

**Table 2:** Transition temperature ($T_c^{onset}$) of all the compounds with their transition width ($\Delta T_c$), normal state resistivity ($\rho_{300}$) and Residual Resistivity Ratio (RRR) are compared.

| Compound | $T_c^{onset}$ (K) | $\Delta T_c$ (K) | $\rho_{300}$ (mΩ-cm) | RRR |
|---|---|---|---|---|
| NdFeAsO$_{0.8}$F$_{0.2}$ | 51.8 | 5 | 2.58 | 4.64 |
| NdFeAsO$_{0.8}$F$_{0.2}$ + 5% Ag | 48.8 | 5 | 2.18 | 3.37 |
| NdFeAsO$_{0.8}$F$_{0.2}$ + 10% Ag | 49.3 | 4.5 | 1.92 | 3.27 |
| NdFeAsO$_{0.8}$F$_{0.2}$ + 20% Ag | 47.2 | 3.8 | 1.58 | 2.78 |

Figure 1

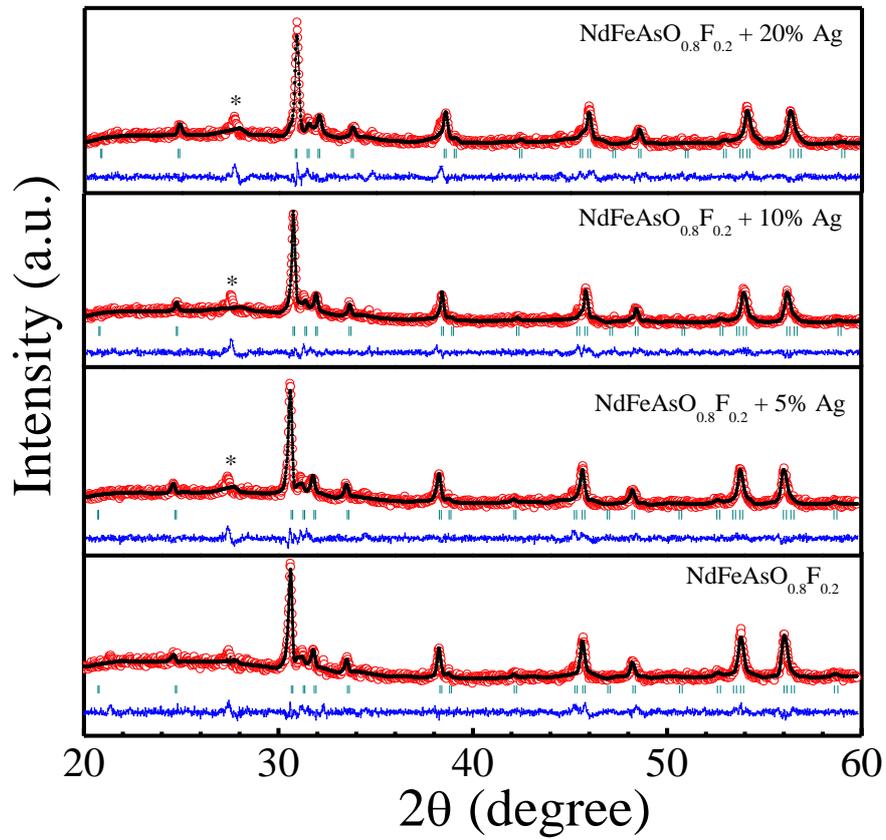

Figure 2

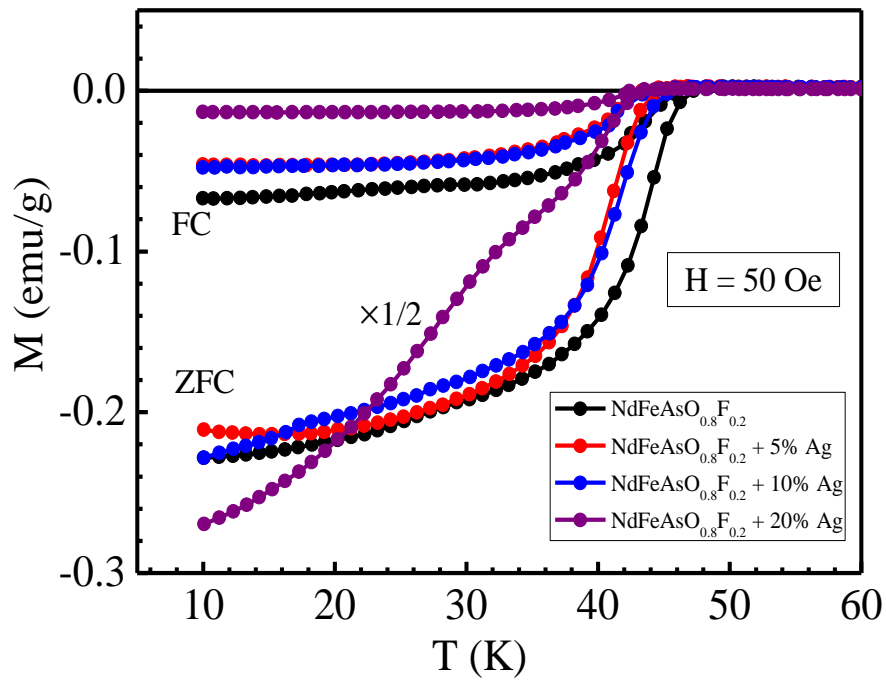

Figure 3

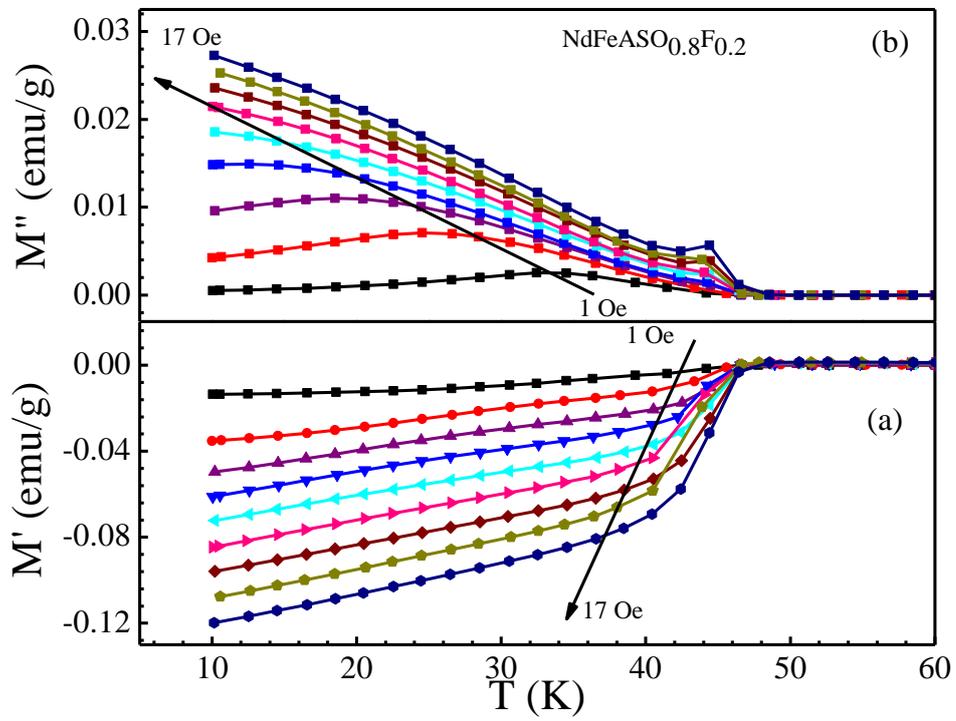

Figure 4

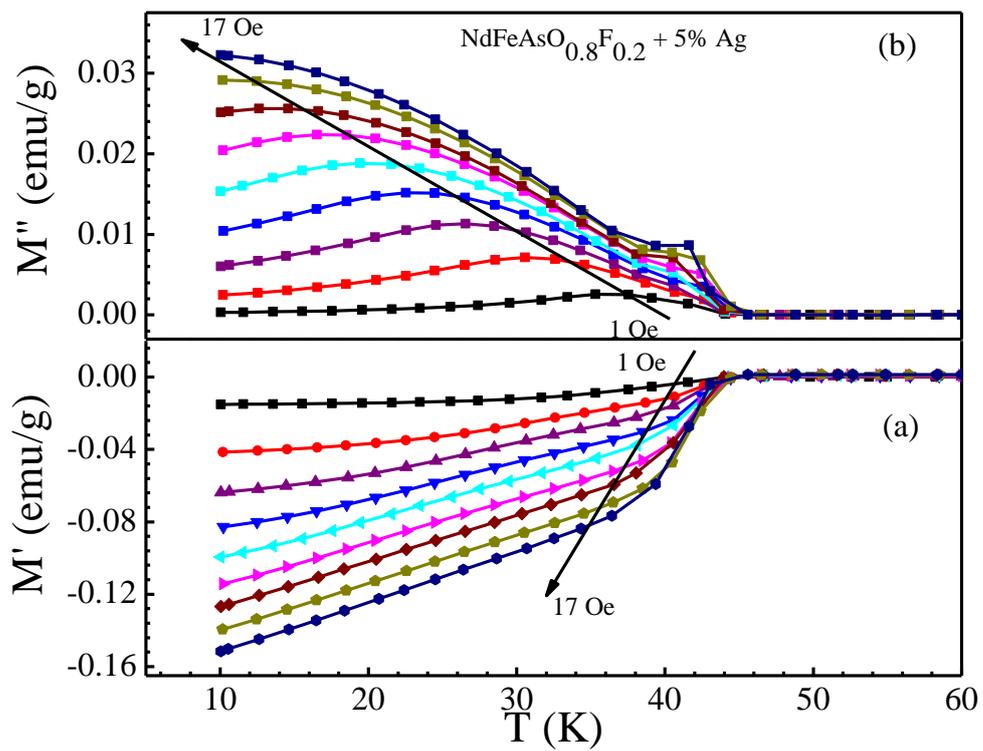

Figure 5

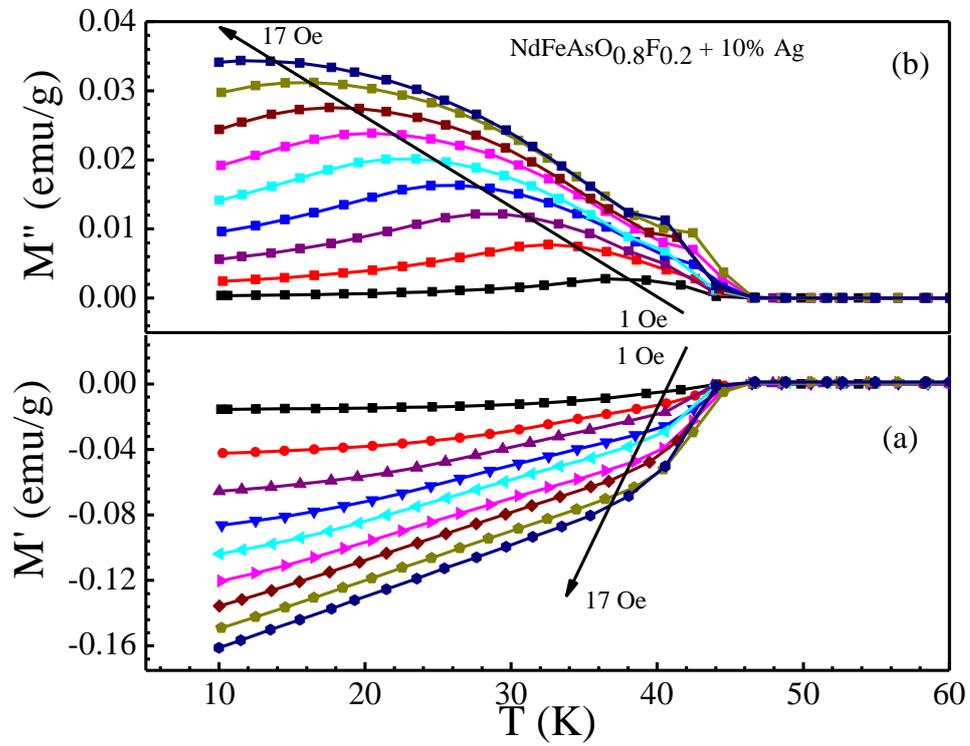

Figure 6

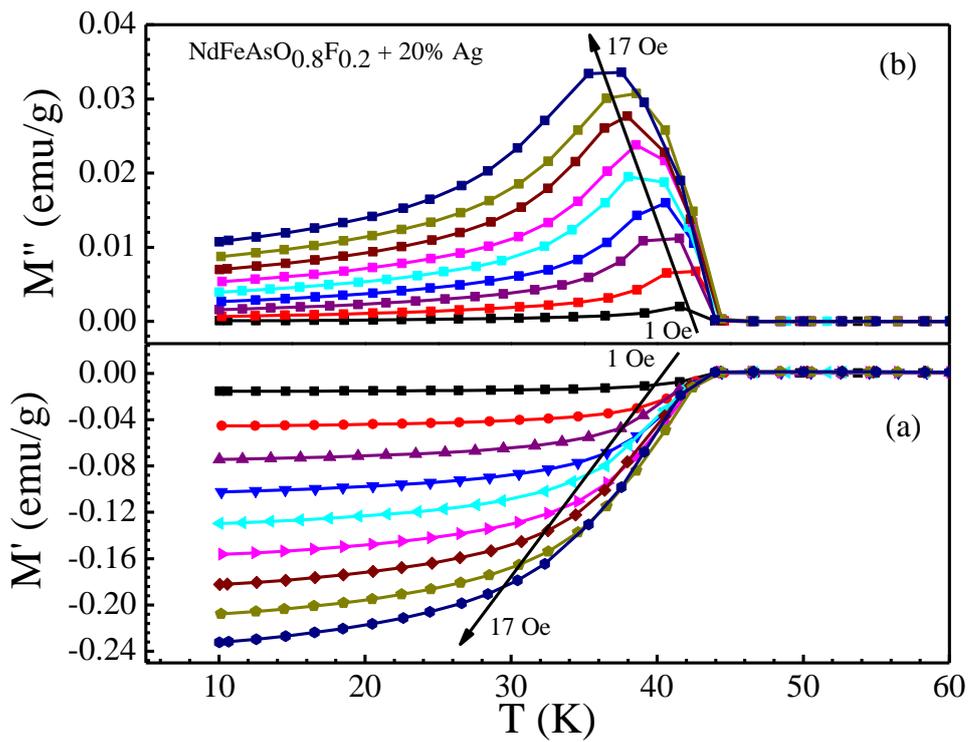

Figure 7

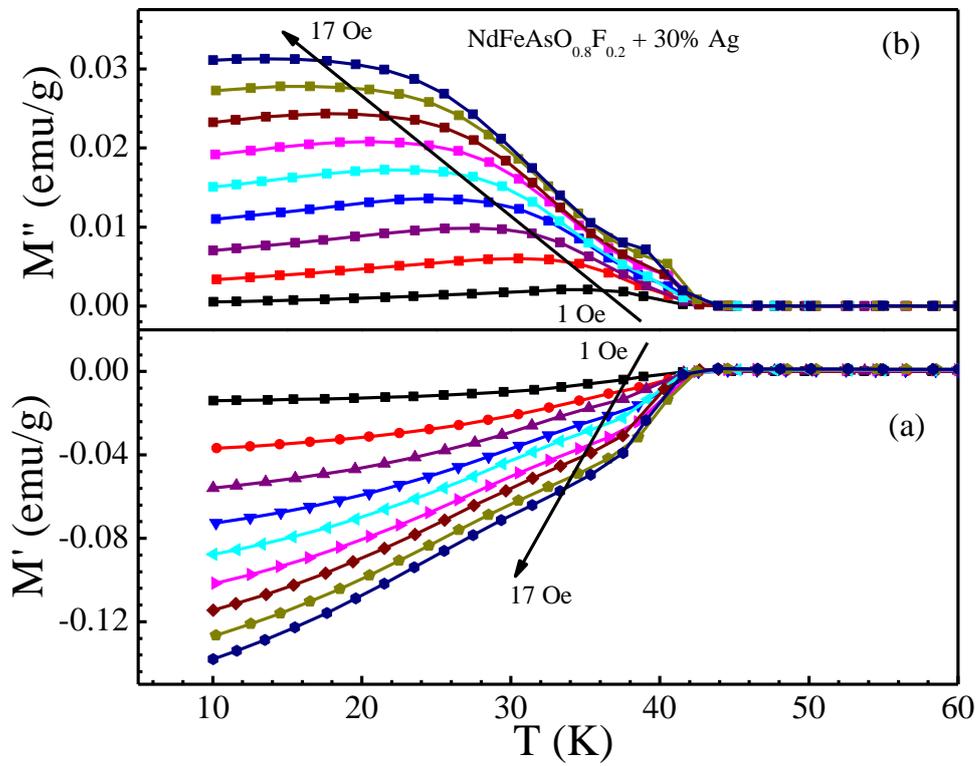

Figure 8

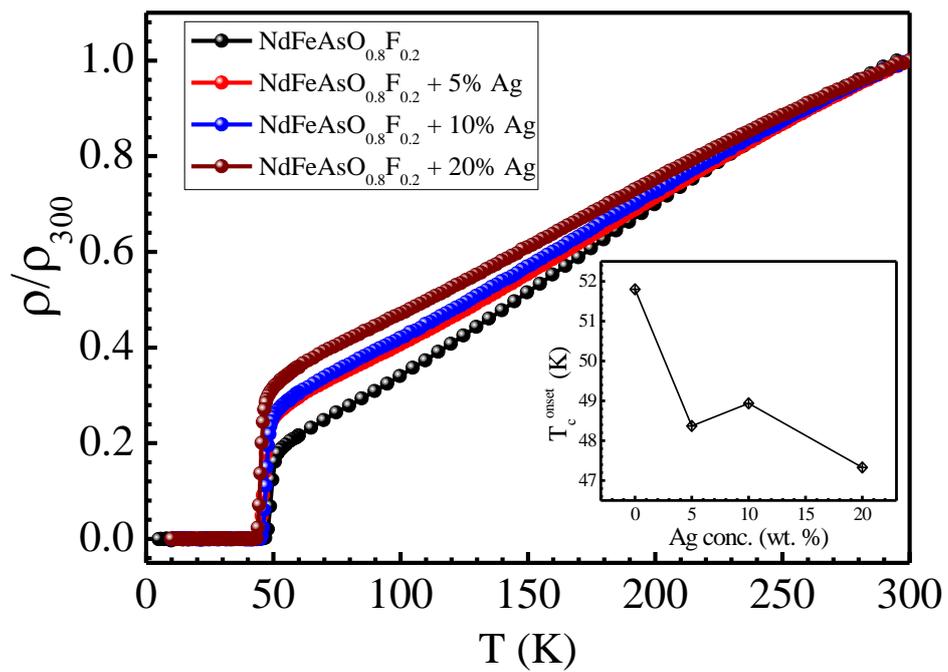

Figure 9

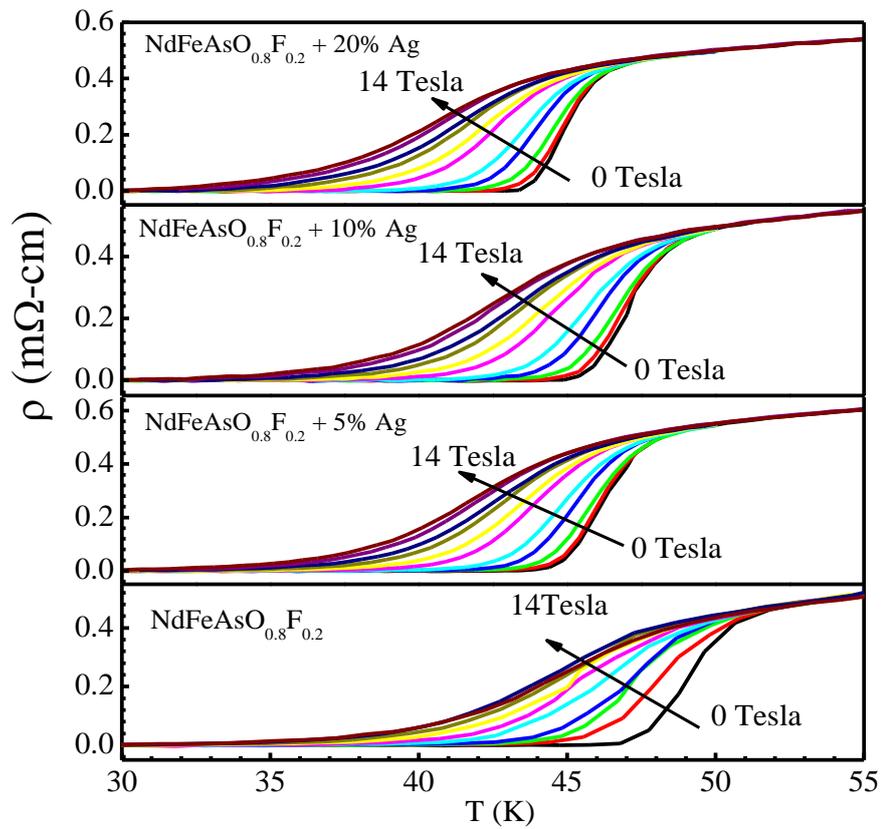

Figure 10

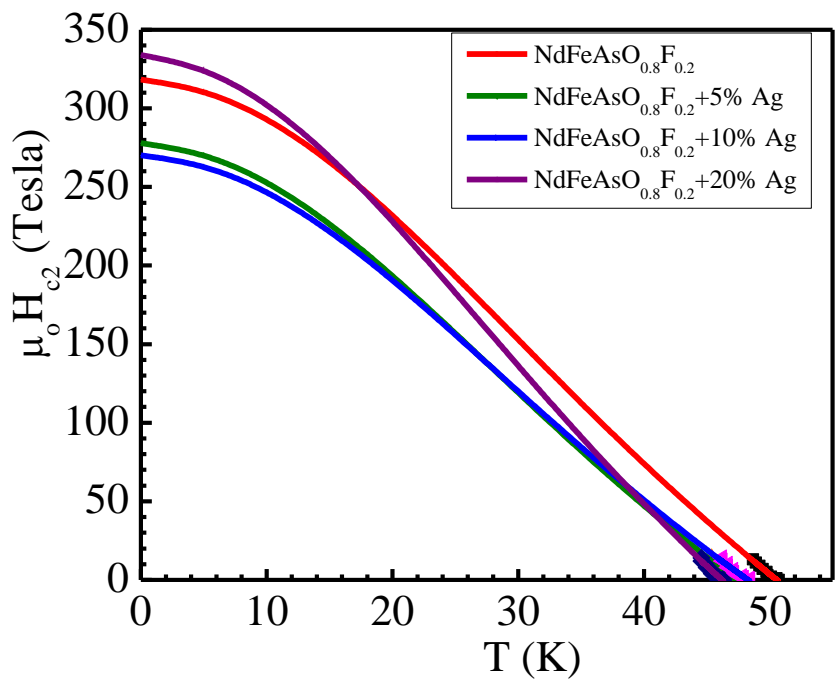

Figure 11

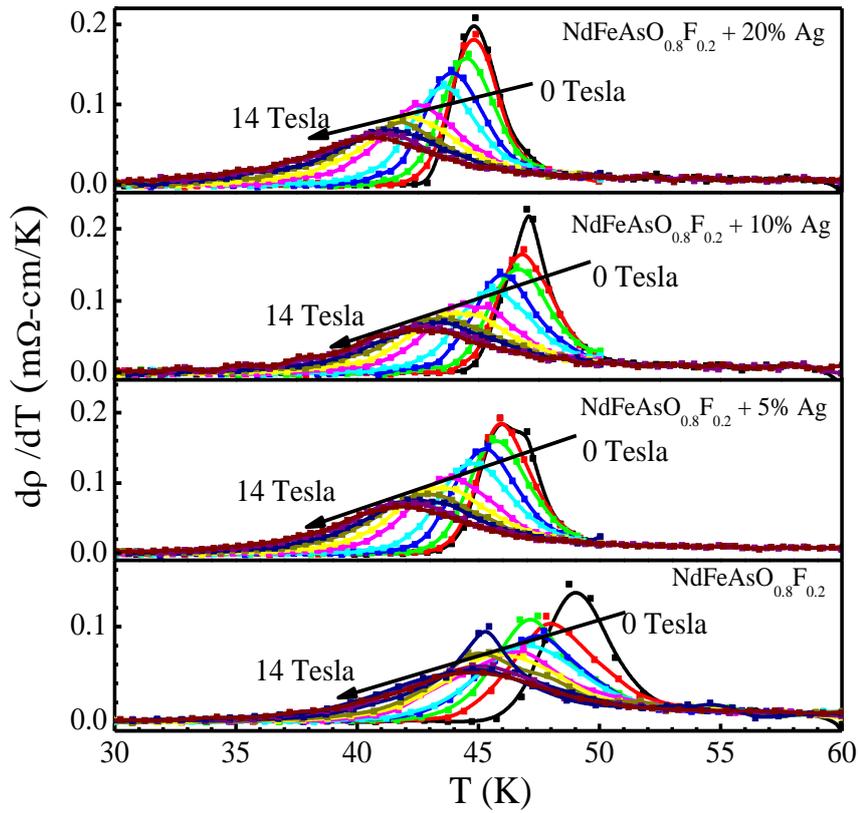

Figure 12

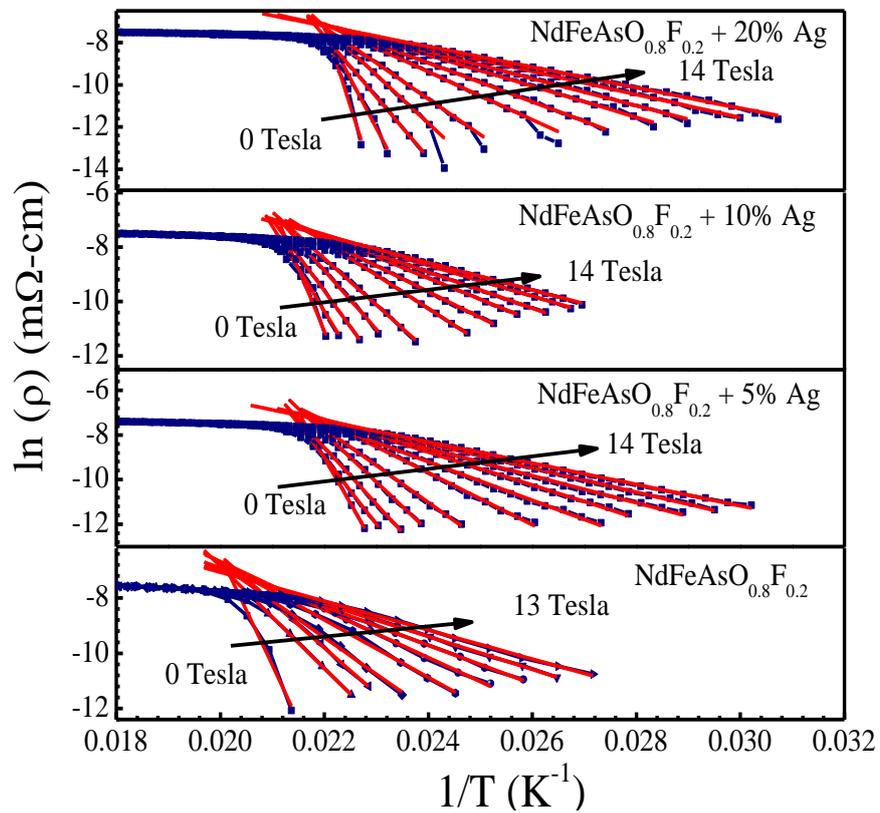

Figure 13

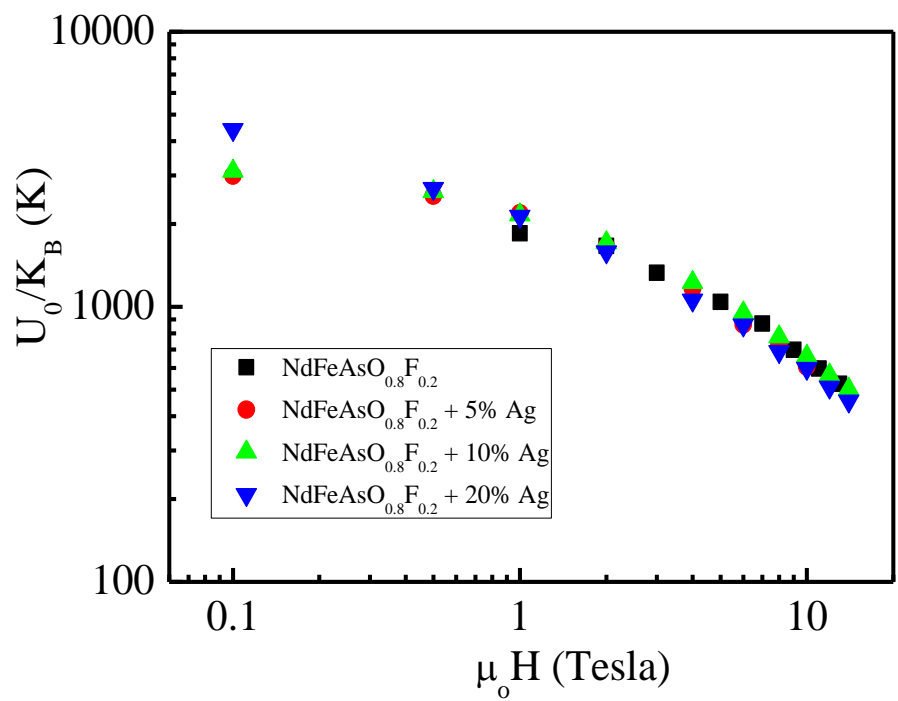

Figure 14

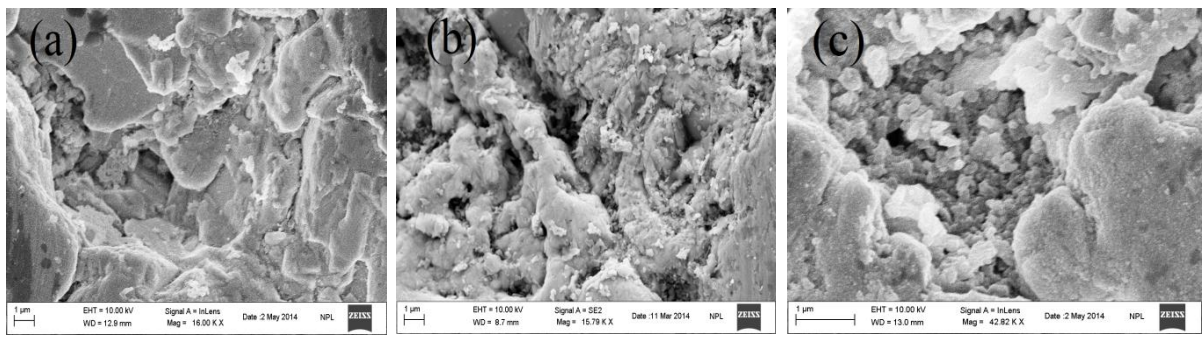